# Variable Mass and the Noisy Feynman Propagator in Scalar Fields

Allan Tameshtit

*allan.tameshtit@utoronto.ca*

We utilize a mass independent Klein-Gordon equation that is first order in a variable that plays the role of time, the approach taken in parametric time formulations. Using concepts from semigroup evolution, we examine the sign of a noisy Feynman propagator in a quantum field theory, namely, scalar electrodynamics.

## I.    Introduction

Special relativity motivates us to treat all components of a four-vector as symmetrically as possible. Nevertheless, the zeroth component often plays a distinguished role, not only having a different sign in the signature, but, when this component is time, being the unique parameter in the group of unitary evolution operators in quantum mechanics. In an effort to relieve time of any special role, we often seek scalars to act as proxies.

The obvious scalar for this purpose is proper time, $t_{\text{pr}}$, which is the time as measured in the rest frame of a particle and which arises in the invariant inner product $x^2 \equiv t^2 - \mathbf{x}^2 = t_{\text{pr}}^2$, where throughout this paper we will employ natural units.

Another such scalar, an invariant parameter conjugate to the mass or square of the mass, is parametric or historical time [1], [2], [3], [4]. After replacing the mass squared with $i\frac{\partial}{\partial \tau_E}$ in the Klein-Gordon equation, Enatsu [5] analyzed the following equation involving the parametric time $\tau_E$,

$$i\frac{\partial \varphi^{(+)}(x,\tau)}{\partial \tau_E} = \left(\frac{\partial^2}{\partial t^2} - \nabla^2\right)\varphi^{(+)}(x,\tau_E) \quad (1)$$

obtaining a solution

$$\varphi^{(+)}(x,\tau_E) = \frac{r_0}{(2\pi)^{1/2}} \int \varphi^{(+)}(x,m^2)\theta(m^2)e^{im^2\tau_E}\,dm^2 \quad (2)$$

with inverse Fourier transform

$$\varphi^{(+)}(x,m^2)\theta(m^2) = \frac{1}{(2\pi)^{1/2}r_0} \int \varphi^{(+)}(x,\tau)e^{-im^2\tau_E}\,d\tau_E \quad (3)$$

provided $\varphi^{(+)}(x,m^2)$ is a solution of the Klein-Gordon equation.

He went on to write the field in terms of ladder operators

$$\varphi^{(+)}(x,m^2) = \frac{1}{(2\pi)^{3/2}} \int e^{-i(k^0 x^0 - \mathbf{k}^2)}\,\delta\!\left(k^{0\,2} - \mathbf{k}^2 - m^2\right)\theta(k_0)a(k,m^2)\,d^4k \quad (4)$$

where

$$[a(k,m^2), a^\dagger(k',m'^2)] = 2k_0\delta(\mathbf{k}-\mathbf{k}')\delta\!\left(r_0^2(m^2 - m'^2)\right) \quad (5)$$

with $k_0 = \sqrt{\mathbf{k}^2 + m^2}$, and $r_0$ the classical electron or Thompson radius [5].

Earlier, by replacing the mass squared with $i\frac{\partial}{\partial u}$ in the minimal coupling Klein-Gordon equation, Feynman [6], based on work by Fock [7], obtained and analyzed the following equation

$$i\frac{\partial \psi_{\text{Fey}}(x,u)}{\partial u} = \frac{1}{2}(i\partial_\mu - eA_\mu)(i\partial^\mu - eA^\mu)\psi_{\text{Fey}}(x,u) \quad (6)$$

from which a solution of the Klein-Gordon equation was obtained via the inverse Fourier transform $\Psi_{\text{Fey}}(x) = \int_{-\infty}^{\infty} \exp\!\left(-\frac{1}{2}im^2 u\right)\psi_{\text{Fey}}(x,u)\,du$.

Parametric time in quantum physics is further discussed in the reviews [8], [9] and monograph [10], and in the references cited therein. Eq. (2) makes clear that the foregoing and other similar theories treat mass as a variable [8], [4], which is the point of view we will often take in this paper.

A scalar $\tau$ will be introduced below as part of the Laplace transform pair $(\xi, \tau)$ and exploited below in quantum field theory, where $\xi = m^2$ is a Lorentz invariant on account of

$$k^2 \equiv k_0^2 - \mathbf{k}^2 = \xi \qquad (7)$$

For a system of several identical particles, the mass has the advantage of not only being an invariant, but of being the same for each particle, which cannot be said about proper times.

Eq. (7) is usually the launching point towards relativistic quantum mechanics of a spinless bosonic particle of mass $m_\mathrm{p}$ [11], the subscript "p" standing for a *p*articular mass of the particle under study. Making the usual replacement $(k^0, \mathbf{k}) \to \left(i\frac{\partial}{\partial t}, -i\boldsymbol{\nabla}\right)$ suggests the Klein-Gordon equation

$$[(i\partial)^2 - \xi_\mathrm{p}]\phi_\mathrm{KG} = 0 \qquad (8)$$

and

$$[(i\partial)^{*2} - \xi_\mathrm{p}]\phi^*_\mathrm{KG} = 0 \qquad (9)$$

with $\partial^2 = \partial^\mu \partial_\mu$ and $\xi_\mathrm{p} = m_\mathrm{p}^2$ in natural units (or $\left(\frac{\hbar}{m_\mathrm{p} c}\right)^{-2}$, the inverse square of the reduced Compton wavelength, upon inserting implied constants).

We may also reverse the steps and derive Eq. (7) from the Klein-Gordon equation, which is more in keeping with the approach we take below to arrive at the aforementioned scalar $\tau$. Solving the Klein-Gordon equation using a separation of variables approach leads to four separation constants $k^\mu$ and solutions that are superpositions of

$$\gamma_{KG}(k)e^{\pm ik_0x_0}e^{\pm ik_1x_1}e^{\pm ik_2x_2}e^{\pm ik_3x_3} \tag{10}$$

where $\gamma_{KG}(k)$ is relativistically invariant and the mass-shell condition, $k^2 = \xi_p$, holds. This last condition is Eq. (7) when we take $\xi = \xi_p$.

It is instructive for our purposes to briefly outline the further steps required to reach quantum field theory [11], [12] and at the same time introduce some notation that we will need. To satisfy Eq. (7), one typically considers the energy $k^0$ to be a dependent function of the momentum $\mathbf{k}$, resulting in the dispersion relation

$$k_0 = \sqrt{\mathbf{k}^2 + \xi_p} \equiv \omega_{\mathbf{k},\xi_p} \tag{11}$$

and the two solutions $\gamma_{KG}^{(\pm)}(\omega_{\mathbf{k},\xi_p}, \mathbf{k}) e^{\mp i(\omega_{\mathbf{k},\xi_p}x^0 - \mathbf{k}\cdot\mathbf{x})}$. Two issues then arise. First, singling out $k_0$ (and $x^0$) in this manner is not an invariant process in the sense that generically $\omega_{\mathbf{k}} \neq \omega_{\mathbf{k}'}$ (and, of course, $x^0 \neq x'^0$) with a Lorentz transformation to primed variables. Second, when forming a superposition, an invariant measure is $d^4k$, but the mass-shell condition implies the four variables $k^\mu$ are not linearly independent, leading to questions about how the integration with respect to these four variables should be performed. The tonic we need is to take

$$\gamma_{KG}^{(\pm)}(k) = \frac{1}{(2\pi)^3}\delta(k^2 - \xi_p)\theta(k^0)\Phi_{KG}^{(\pm)}(k), \tag{12}$$

where $\Phi_{KG}^{(\pm)}(k)$ are invariant functions [12]. The advantage of this expression is that the resulting invariant integral can be performed as if all four variables were independent, letting the function $\delta(k^2 - \xi_p)\theta(k^0)$ do the work of restricting the integration to the mass-shell with positive energy:

$$\phi_{KG}^{(+)}(x) = \frac{1}{(2\pi)^3}\int e^{-ikx}\Phi_{KG}^{(+)}(k)\delta(k^2 - \xi_p)\theta(k^0)\, d^4k \tag{13}$$

$$= \int e^{-i(\omega_{\mathbf{k},\xi_p}t - \mathbf{k}\cdot\mathbf{x})} \Phi_{KG}^{(+)}(\omega_{\mathbf{k},\xi_p}, \mathbf{k}) \frac{1}{2(2\pi)^3 \omega_{\mathbf{k},\xi_p}} d^3k \qquad (14)$$

where we have used $\delta(k^2 - \xi_p)\theta(k^0) = \frac{\delta(k^0 - \omega_{\mathbf{k},\xi_p})}{2\omega_{\mathbf{k},\xi_p}}$ with $\omega_{\mathbf{k},\xi_p} = \sqrt{\mathbf{k}^2 + \xi_p}$.

The states $f_{\mathbf{k}}(x) = [(2\pi)^3 2\omega_{\mathbf{k},\xi_p}]^{-1/2} e^{-i(\omega_{\mathbf{k},\xi_p}t - \mathbf{k}\cdot\mathbf{x})}$ and $\phi_{KG}^{(+)}(x)$ are called "positive energy solutions," the latter superpositions, unlike $f_{\mathbf{k}}(x)$, having finite norm. These positive energy states can be made into a Hilbert space on the mass-shell [12]. Using the specific inner products $\langle \mathbf{k}'|\mathbf{k}\rangle = (2\pi)^3 2\omega_{\mathbf{k}} \delta(\mathbf{k} - \mathbf{k}')$ [11] and $\langle x|\mathbf{k}\rangle = e^{-i(\omega_k x^0 - \mathbf{k}\cdot\mathbf{x})}$, and two alternate resolutions of the identity

$$\int d^3x [|0,\mathbf{x}\rangle\langle 0,\mathbf{x}|, \hat{k}^0]_+ = \int d^3k \, |\mathbf{k}\rangle \frac{1}{(2\pi)^3 2\omega_{\mathbf{k}}} \langle \mathbf{k}| = 1_{KG}, \qquad (15)$$

the general inner product $\langle \phi_{KG}^{(+)}|\psi_{KG}^{(+)}\rangle$, when both $\langle x|\phi_{KG}^{(+)}\rangle = \phi_{KG}^{(+)}(x)$ and $\psi_{KG}^{(+)}(x)$ are positive energy solutions of the Klein-Gordon equation, can be calculated. In the respective position and momentum representations,

$$\langle \phi_{KG}^{(+)}|\psi_{KG}^{(+)}\rangle = \int \phi_{KG}^{(+)*}(x) i \overleftrightarrow{\partial_0} \psi_{KG}^{(+)}(x) d^3x \qquad (16)$$

$$= \int \Phi_{KG}^{(+)*}(\omega_{\mathbf{k}}, \mathbf{k}) \frac{1}{(2\pi)^3 2\omega_{\mathbf{k}}} \Psi_{KG}^{(+)}(\omega_{\mathbf{k}}, \mathbf{k}) d^3k \qquad (17)$$

where $\Phi_{KG}^{(+)}(\omega_{\mathbf{k}}, \mathbf{k}) = \langle \mathbf{k}|\phi_{KG}^{(+)}\rangle$ and $A \overleftrightarrow{\partial_0} B \equiv A \frac{\partial B}{\partial t} - \frac{\partial A}{\partial t} B$ [12]. Eq. (17) is the analogue of the Parseval-Plancherel identity in the Klein-Gordon inner product space.

The set $\{f_{\mathbf{k}}(x)\}$ is orthonormal, $\int f_{\mathbf{k}}^*(x) i \overleftrightarrow{\partial_0} f_{\mathbf{k}'}(x) d^3x = \delta(\mathbf{k} - \mathbf{k}')$. One can invert Eq. (14) by using the following inner product

$$\Phi_{KG}^{(+)}(\omega_{\mathbf{k},\xi_p}, \mathbf{k}) = \int e^{i(\omega_{\mathbf{k},\xi_p}t - \mathbf{k}\cdot\mathbf{x})} i \overleftrightarrow{\partial_0} \phi_{KG}^{(+)}(x) d^3x$$

$$= \int \left[(2\pi)^3 2\omega_{\mathbf{k},\xi_p}\right]^{1/2} f_{\mathbf{k}}^*(x) i \overleftrightarrow{\partial_0} \phi_{KG}^{(+)}(x) \, d^3x$$

$$= \langle \mathbf{k} | \phi_{KG}^{(+)} \rangle. \tag{18}$$

To transition from relativistic quantum mechanics---where $\Phi_{KG}^{(+)}(k)$ is an invariant function that need only yield a finite norm but is otherwise arbitrary---to quantum field theory, we replace $\Phi_{KG}^{(+)}(\omega_{\mathbf{k}}, \mathbf{k})$ by an operator $a_{KG}(\mathbf{k})$ with commutator [11]

$$[a_{KG}(\mathbf{k}), a_{KG}^\dagger(\mathbf{k}')] = (2\pi)^3 2\omega_{\mathbf{k},\xi_p} \delta(\mathbf{k} - \mathbf{k}'). \tag{19}$$

The creation operator creates a particle in the state $e^{-i(\omega_{\mathbf{k},\xi_p} t - \mathbf{k}\cdot\mathbf{x})}$ from the vacuum, $\langle x | a_{KG}^\dagger(\mathbf{k}) | 0 \rangle = \langle x | \mathbf{k} \rangle_{KG} = e^{-i(\omega_{\mathbf{k},\xi_p} t - \mathbf{k}\cdot\mathbf{x})}$. To allow for charged particles, we also introduce a creation operator $b_{KG}^\dagger$ that creates an antiparticle, and also satisfies the commutation relation

$$[b_{KG}(\mathbf{k}), b_{KG}^\dagger(\mathbf{k}')] = (2\pi)^3 2\omega_{\mathbf{k},\xi_p} \delta(\mathbf{k} - \mathbf{k}'). \tag{20}$$

The field operator is given by the particular operator solution of the Klein-Gordon equation [11],

$$\phi_{KG}(x) = \int \frac{d^4k}{(2\pi)^3} \left[a_{KG}(\mathbf{k}) e^{-ikx} + b_{KG}^\dagger(\mathbf{k}) e^{ikx}\right] \delta(k^2 - \xi_p) \theta(k^0) \tag{21}$$

$$= \int \frac{d^3k}{(2\pi)^3 2\omega_{\mathbf{k},\xi_p}} \left[a_{KG}(\mathbf{k}) e^{-i(\omega_{\mathbf{k},\xi_p} t - \mathbf{k}\cdot\mathbf{x})} + b_{KG}^\dagger(\mathbf{k}) e^{i(\omega_{\mathbf{k},\xi_p} t - \mathbf{k}\cdot\mathbf{x})}\right]. \tag{22}$$

The commutators Eq. (19) and Eq. (20) ensure that the equal time commutator $[\phi_{KG}(x^0, \mathbf{x}), \pi_{KG}(x^0, \mathbf{y})] = i\delta(\mathbf{x} - \mathbf{y})$, where $\pi_{KG}(x) = \dot{\phi}_{KG}^\dagger(x)$. The quantity $\pi_{KG}(x)$ is the momentum field conjugate to $\phi_{KG}(x)$.

Unlike $x^2$, which can take any value in spacetime (tantamount to picking an arbitrary origin for our reference frame), we usually restrict $k^2$ to have timelike values, and more specifically, the value $m_p^2$, where $m_p$ is the particular mass of the particle under consideration. Since we wish to use a variable $\tau$ conjugate to $\xi = m^2$ that, like the proper time, can take on many values, we relax--at least initially--this condition and, as in [4], allow $\xi$ to also assume various values. In quantum field theory, we utilize ladder operators that are associated with varying masses, which satisfy

$$[a(k), a^\dagger(k)] = [b(k), b^\dagger(k)] = \delta^4(k - k'), \tag{23}$$

a commutator that appears in [3]. Whereas the particle created from the vacuum with $a_{KG}^\dagger(\mathbf{k})$ has a fixed energy once the momentum $\mathbf{k}$ is specified (in line with the dispersion relation), the operator $a^\dagger(k)$ treats energy and $\mathbf{k}$ independently by permitting the creation of particles with arbitrary 4-momentum, $\langle x|a^\dagger(k)|0\rangle = \langle x|k\rangle = e^{-ikx}$.

To connect with conventional approaches, we may confine a superposition to the $\xi_p$ mass-shell to yield the following operator solution of the Klein-Gordon equation found in the literature [13], [5], [3]:

$$\phi(x, \xi_p) = \int \frac{d^4k}{(2\pi)^{3/2}} [a(k)e^{-ikx} + b^\dagger(k)e^{ikx}]\delta(k^2 - \xi_p)\theta(k^0) \tag{24}$$

Equations (21) and (24) are similar except for the different creation (and annihilation) operators that in the former create a specific massive particle, and in the latter create particles of varying masses (or energies).

In quantum field theory, the foregoing considerations will lead us to the mass independent Klein-Gordon equation (MIKE) in which an invariant parameter $\tau$, one of a Laplace transform pair $(\xi, \tau)$, plays the role of time, like the parametric time approaches described above. To obtain the "initial condition" at $\tau = 0$, as well as other quantities appearing below, we integrate quantum fields over $\xi$, as in the general expression $\int_0^\infty \rho(\xi) D(\xi, x)\, d\xi$, where $D(\xi, x)$ is a two-point correlator and $\rho(\xi)$ is a mass spectral density, and as in the specific expression $i \int_0^\infty \rho(\xi) \frac{1}{p^2 - \xi + i\varepsilon}\, d\xi$, both known as KL spectral representations [14], [15].

Among the benefits of freeing up physical time in this way is that MIKE is first order in $\tau$ and we can express evolution with respect to $\tau$ in a form that is similar to completely positive maps in semigroup formulations [16], allowing certain sign properties of two-point correlators to become more transparent.

Below, starting in Section II, we will look at some of the consequences of thinking of the mass as a variable rather than a fixed parameter, relating expressions that involve the conventional ladder operators $a_{\text{KG}}$ and $b_{\text{KG}}$ to those that involve the operators $a$ and $b$. In Section III, we introduce MIKE by Laplace transforming variables from $\xi$ to $\tau$, and also examine a connection between some two-point correlators and convolutions. In Section IV, we look at coupled systems and in particular, scalar electrodynamics. We exploit some of the advantages of working with $\tau$ in Section V, where we examine a noisy Feynman propagator by taking advantage of the positive dynamical map formalism. The Appendix lists some relations involving positive cone solutions.

## II.  Mass as a Variable

The fields $\phi_{\text{KG}}(x, \xi_{\text{p}})$ and $\phi(x, \xi_{\text{p}})$ are both operator solutions of the same Klein-Gordon equation---the delta function confining the integrals in Eq. (21) and Eq. (24) to the same $\xi_{\text{p}}$ mass shell---albeit unequal solutions because of the disparate ladder operators appearing in their defining sums. The states $e^{-i(\omega_{\mathbf{k},\xi_p}t - \mathbf{k}\cdot\mathbf{x})}$, too, are wave function solutions of this same Klein-Gordon equation with fixed $\xi_p$.

On the other hand, the wave functions $e^{-ikx} = \langle x|a^\dagger(k)|0\rangle$, with timelike but otherwise varying $k^2$, are solutions of respective members of a family of Klein-Gordon equations, each having a different mass satisfying $k^2 = \xi$. Rather than framing our discussion in terms of an operator $a$, as appears in Eq. (24), whose adjoint can create states off the $\xi_p$ mass-shell, it is more in keeping with the present formalism to say that $a^\dagger$ can create states on various mass-shells inside the positive cone $(k^0, k^2 > 0)$.

Operator solutions of $[(i\partial)^2 - \xi]\phi = 0$ include superpositions of $e^{\pm i(\omega_{\mathbf{k},\xi}x^0 - \mathbf{k}\cdot\mathbf{x})}$ over $\mathbf{k}$,

$$\phi(x,\xi) = \int \hat{\gamma}_+(\omega_{\mathbf{k},\xi},\mathbf{k}) e^{-i(\omega_{\mathbf{k},\xi}x^0 - \mathbf{k}\cdot\mathbf{x})} d^3k + \int \hat{\gamma}_-(\omega_{\mathbf{k},\xi},\mathbf{k}) e^{i(\omega_{\mathbf{k},\xi}x^0 - \mathbf{k}\cdot\mathbf{x})} d^3k, \tag{25}$$

where $\hat{\gamma}_\pm(\omega_{\mathbf{k},\xi},\mathbf{k})$ are operator-valued functions that depend on the ladder operators and $\omega_{\mathbf{k},\xi} = \sqrt{\mathbf{k}^2 + \xi}$. In the following, we may distinguish a particular mass $m_p = \sqrt{\xi_p}$ and take as solutions

$$\phi(x,\xi_p) = \int\int_0^\infty a(\omega_{\mathbf{k},\xi},\mathbf{k}) e^{-i(\omega_{\mathbf{k},\xi}t - \mathbf{k}\cdot\mathbf{x})} \delta(\xi - \xi_p) \frac{1}{(2\pi)^{3/2} 2\omega_{\mathbf{k},\xi}} d\xi d^3k + \int\int_0^\infty b^\dagger(\omega_{\mathbf{k},\xi},\mathbf{k}) e^{i(\omega_{\mathbf{k},\xi}t - \mathbf{k}\cdot\mathbf{x})} \delta(\xi - \xi_p) \frac{1}{(2\pi)^{3/2} 2\omega_{\mathbf{k},\xi}} d\xi d^3k. \tag{26}$$

Eq. (25) and Eq. (26) are equivalent if we take $\hat{\gamma}_+(\omega_{\mathbf{k},\xi},\mathbf{k}) = \frac{1}{(2\pi)^{3/2} 2\omega_{\mathbf{k},\xi}} a(\omega_{\mathbf{k},\xi},\mathbf{k})$, $\hat{\gamma}_-(\omega_{\mathbf{k},\xi},\mathbf{k}) = \frac{1}{(2\pi)^{3/2} 2\omega_{\mathbf{k},\xi}} b^\dagger(\omega_{\mathbf{k},\xi},\mathbf{k})$ and set $\xi = \xi_p$.

Finally, a change of variables $k^0 = \sqrt{\xi + \mathbf{k}^2}$ leads to

$$\phi(x,\xi_p) = \frac{1}{(2\pi)^{3/2}} \int a(k) e^{-ikx} \delta(k^2 - \xi_p) \theta(k^0) \, d^4k$$
$$+ \frac{1}{(2\pi)^{3/2}} \int b^\dagger(k) e^{ikx} \delta(k^2 - \xi_p) \theta(k^0) \, d^4k, \tag{27}$$

which is the same expression as Eq. (24).

Eq. (27) is seen to be a hybrid of a positive energy solution with the sum restricted to a particular mass-shell, and a positive cone solution with ladder operators that can destroy and create particles having four-momenta satisfying $k^2 > 0$ and $k^0 > 0$. The single particle manifold of standard Klein-Gordon theory obtained from the set of positive energy solutions having a fixed $\xi_p > 0$, $\left\{ e^{-i(\omega_{k,\xi_p} t - \mathbf{k} \cdot \mathbf{x})} | \omega_{k,\xi_p} = \sqrt{\mathbf{k}^2 + \xi_p} \right\}$, has an associated identity operator

$$\frac{1}{(2\pi)^3} \int d^3k \, |\omega_{k,\xi_p}, \mathbf{k}\rangle \frac{1}{2\omega_{k,\xi_p}} \langle \omega_{k,\xi_p}, \mathbf{k}| = 1_{KG} \tag{28}$$

whereas the manifold obtained from the set of positive cone (PC) solutions having variable $\xi$,

$$\left\{ e^{-ikx} | k^0, k^2 > 0 \right\} = \left\{ e^{-i(\omega_{k,\xi} t - \mathbf{k} \cdot \mathbf{x})} | \omega_{k,\xi} = \sqrt{\mathbf{k}^2 + \xi} \text{ and } \xi > 0 \right\}, \tag{29}$$

has an associated identity operator for single particle states,

$$\int_0^\infty d\xi \int d^3k \, |\omega_{k,\xi}, \mathbf{k}\rangle \frac{1}{2\omega_{k,\xi}} \langle \omega_{k,\xi}, \mathbf{k}| = 1_{PC}. \tag{30}$$

When using the former identity operator, the conventional inner product is to be used, $\langle \mathbf{k} | \mathbf{k}' \rangle = (2\pi)^3 2\omega_k \delta(\mathbf{k} - \mathbf{k}')$, whereas when using the latter, the applicable inner product is $\langle \omega_{k,\xi}, \mathbf{k} | \omega_{k',\xi'}, \mathbf{k}' \rangle = \delta(\omega_{k,\xi} - \omega_{k',\xi'}) \delta(\mathbf{k} - \mathbf{k}')$. Comparing the two resolutions of the identity, we see that the manifold of positive cone solutions is obtained by adding individual mass-shells over all possible values of $\xi$. Positive cone solutions are further discussed in the Appendix.

Neglecting the contributions from the ground state energies, it is reasonable to take for the Hamiltonian

$$H = \int k^0 [a^\dagger(k)a(k) + b^\dagger(k)b(k)]\theta(k^2)\theta(k^0)\, d^4k \tag{31}$$

since this operator adds up the number of quanta in an occupation state multiplied by the corresponding energy in the positive cone ($k^2, k^0 > 0$) to yield the total energy. This expression is similar to the conventional one for the Hamiltonian [11], [12],

$$\frac{1}{(2\pi)^3} \int k^0 [a_{KG}^\dagger(\mathbf{k})a_{KG}(\mathbf{k}) + b_{KG}^\dagger(\mathbf{k})b_{KG}(\mathbf{k})]\delta(k^2 - \xi_p)\theta(k^0)d^4k \tag{32}$$

except that the latter is restricted to a sum on the positive energy mass-shell, unlike the integral in Eq. (31), which is over the region inside the positive cone. We note that Eq. (31) for the Hamiltonian is independent of mass. If we change variables according to $k^0 = \sqrt{\mathbf{k}^2 + \xi}$ with $\xi > 0$, we can use Eq. (23) and Eq. (31) to confirm that the solution of $\dot{a}(t) = -i[a(t), H]$ in the positive cone is $a(t) = a(0)e^{-i\omega_{\mathbf{k},\xi}t}$ where $\omega_{\mathbf{k},\xi} \equiv \sqrt{\mathbf{k}^2 + \xi}$, as we would expect.

Eq. (31) becomes

$$H = \frac{1}{2}\int_0^\infty d\xi \int d^3k \left[a^\dagger(\omega_{\mathbf{k},\xi}, \mathbf{k})a(\omega_{\mathbf{k},\xi}, \mathbf{k}) + b^\dagger(\omega_{\mathbf{k},\xi}, \mathbf{k})b(\omega_{\mathbf{k},\xi}, \mathbf{k})\right] \tag{33}$$

after a change of variables. The particle, as opposed to the anti-particle, contribution may be written as

$\int_0^\infty d\xi \int d^4k\, \omega_{\mathbf{k},\xi} a^\dagger(\omega_{\mathbf{k},\xi}, \mathbf{k})a(\omega_{\mathbf{k},\xi}, \mathbf{k})\theta(k^0)\delta(k^2 - \xi),$

which is similar to Eq. 4.21 of [5], but, *inter alia*, with the factor $\omega_{\mathbf{k},\xi}$ in the integrand replaced by $\xi$.

We can arrive at a Hamiltonian that is expressed in terms of the field operator

$$\phi(x,\xi) = \frac{1}{(2\pi)^{3/2}} \int a(k) e^{-ikx} \delta(k^2 - \xi) \theta(k^0)\, d^4k$$
$$+ \frac{1}{(2\pi)^{3/2}} \int b^\dagger(k) e^{ikx} \delta(k^2 - \xi) \theta(k^0)\, d^4k \tag{34}$$

by inverting this last equation (cf. Eq. (24)) to isolate the ladder operators,

$$a(\omega_{\mathbf{k},\xi}, \mathbf{k}) = \int_0^\infty d\xi' \int d^4x \frac{1}{(2\pi)^{5/2}} e^{i(\omega_{\mathbf{k},\xi} x^0 - \mathbf{k}\cdot\mathbf{x})} \phi(x,\xi')$$
$$= \frac{1}{(2\pi)^{3/2}} \int e^{i(\omega_{\mathbf{k},\xi} x^0 - \mathbf{k}\cdot\mathbf{x})} i \overleftrightarrow{\partial_0} \phi(x,\xi)\, d^3x \tag{35}$$

and

$$b(\omega_{\mathbf{k},\xi}, \mathbf{k}) = \frac{1}{(2\pi)^{5/2}} \int_0^\infty d\xi' \int d^4x\, e^{i(\omega_{\mathbf{k},\xi} x^0 - \mathbf{k}\cdot\mathbf{x})} \phi^\dagger(x,\xi')$$
$$= \frac{1}{(2\pi)^{3/2}} \int d^3x\, e^{i(\omega_{\mathbf{k},\xi} x^0 - \mathbf{k}\cdot\mathbf{x})} i \overleftrightarrow{\partial_0} \phi^\dagger(x,\xi) \tag{36}$$

Eq. (33) neglects the infinite contribution from the ground state energies. Reintroducing this contribution is tantamount to replacing $a^\dagger(k)a(k)$ by $\frac{1}{2}[a^\dagger(k)a(k) + a(k)a^\dagger(k)]$ and $b^\dagger(k)b(k)$ by $\frac{1}{2}[b^\dagger(k)b(k) + b(k)b^\dagger(k)]$ in Eq. (33), the result of which we denote by $\overline{H}$. It is this Hamiltonian $\overline{H}$ that we may express in terms of the field operators by replacing the ladder operators with the help of Eqs. (35) and (36) to obtain

$$\overline{H} = \int_0^\infty d\xi \int d^3x\, \dot{\phi}^\dagger(x,\xi)\dot{\phi}(x,\xi) + \int_0^\infty d\xi \int d^3x\, \nabla\phi^\dagger(x,\xi) \cdot \nabla\phi(x,\xi)$$
$$+ \int_0^\infty d\xi \int d^3x\, \xi \phi^\dagger(x,\xi)\phi(x,\xi) \tag{37}$$

where to derive this last equation, boundary terms were neglected in which the fields are evaluated at $\|x\| \to \pm\infty$. Just like in the conventional theory, this Hamiltonian is beset with infinities; for example, $\langle 0|\overline{H}|0\rangle \to \infty$.

The associated Hamiltonian density is
$$\mathcal{H} = \Pi^\dagger \Pi + \nabla\phi^\dagger \cdot \nabla\phi + \xi\phi^\dagger\phi \tag{38}$$
where the field momentum conjugate to $\phi$ is $\Pi = \dot\phi^\dagger$ and $\overline{H} = \int_0^\infty \int \mathcal{H}\, d^3x\, d\xi$.

The dimensions of $\mathcal{H}$ (after multiplying the right-hand side of Eq. (38) by an implied factor $\hbar c$) are those of a linear energy density, energy/length, instead of the conventional energy/length$^3$.

The equal time commutation relation for the fields is
$$[\phi(x^0, x, \xi), \Pi(x^0, y, \xi')] = i\delta(x-y)\delta(\xi-\xi') \tag{39}$$
Keeping in mind that $\dot\phi^\dagger = \Pi$, we may use Eq. (39) and the Heisenberg equation $d\dot\phi/dt = -i[\dot\phi, \overline{H}]$ to recover the Klein-Gordon equation.

Several conventional expressions involving pairs of field operators can be rewritten in terms of the field variables introduced above with variable $\xi$ and commutators $[a(k), a^\dagger(k')] = [b(k), b^\dagger(k')] = \delta^4(k-k')$.

For example, single particle ("+" subscripts) and antiparticle ("-" subscripts) matrix elements of the conventional Hamiltonian can be related to those of the Hamiltonian of Eq. (33) according to
$$\langle \mathbf{k}'_-|\langle \mathbf{k}'_+|H_{\text{KG}}|\mathbf{k}_+\rangle|\mathbf{k}_-\rangle$$
$$= (2\pi)^6 \int_0^\infty d\xi'_+ \int_0^\infty d\xi'_- \langle \omega_{\mathbf{k}'_-,\xi'_-}, \mathbf{k}'_-|\langle \omega_{\mathbf{k}'_+,\xi'_+}, \mathbf{k}'_+|H|\omega_{\mathbf{k}_+,\xi_p}, \mathbf{k}_+\rangle|\omega_{\mathbf{k}_-,\xi_p}, \mathbf{k}_-\rangle \tag{40}$$
$$= ((2\pi)^3 2)^2 \left(\omega_{\mathbf{k}_+,\xi_p} + \omega_{\mathbf{k}_-,\xi_p}\right) \omega_{\mathbf{k}_+,\xi_p} \omega_{\mathbf{k}_-,\xi_p} \delta(\mathbf{k}_+ - \mathbf{k}'_+)\delta(\mathbf{k}_- - \mathbf{k}'_-)$$

Other examples involve the equal-time commutator

$$[\phi_{\text{KG}}(x^0, \mathbf{x}), \pi_{\text{KG}}(x^0, \mathbf{y})] = \int_0^\infty [\phi(x^0, \mathbf{x}, \xi), \pi(x^0, \mathbf{y}, \xi')] \, d\xi' \tag{41}$$

$$= i\delta(\mathbf{x} - \mathbf{y})$$

and ladder commutator

$$[a_{\text{KG}}(\mathbf{k}), a_{\text{KG}}^\dagger(\mathbf{k}')] = (2\pi)^3 \omega_{\mathbf{k},\xi} \int_0^\infty [a(\omega_{\mathbf{k},\xi''}, \mathbf{k}), a^\dagger(\omega_{\mathbf{k}',\xi''}, \mathbf{k}')] \frac{1}{\omega_{\mathbf{k},\xi''}} d\xi'' \tag{42}$$

$$= (2\pi)^3 2\omega_{\mathbf{k},\xi} \delta(\mathbf{k} - \mathbf{k}').$$

We also have

$$\langle 0|[\phi_{\text{KG}}(x), a_{\text{KG}}^\dagger(\mathbf{k})]|0\rangle = (2\pi)^{3/2} \int_0^\infty \langle 0|[\phi(x, \xi'), a^\dagger(\omega_{\mathbf{k},\xi}, \mathbf{k})]|0\rangle d\xi' \tag{43}$$

$$= \langle x|\mathbf{k}\rangle$$

$$= e^{-i(\omega_{\mathbf{k},\xi} x^0 - \mathbf{k}\cdot\mathbf{x})}.$$

Similar to the set $\{f_{\mathbf{k}}(x)\}$, we have the orthonormal set

$\left\{g_{\mathbf{k},\xi}(x) = \frac{1}{(2\pi)^2 \sqrt{2\omega_{\mathbf{k},\xi}}} e^{-i(\omega_{\mathbf{k},\xi} t - \mathbf{k}\cdot\mathbf{x})}\right\}$ satisfying

$$\int g_{\mathbf{k},\xi}^*(x) g_{\mathbf{k}',\xi'}(x) d^4 x = \delta(\xi - \xi') \delta(\mathbf{k} - \mathbf{k}') \tag{44}$$

Then, analogous to the inversion relation [11]

$a_{\text{KG}}(\mathbf{k}) = [(2\pi)^3 2\omega_{\mathbf{k}}]^{1/2} \int f_{\mathbf{k}}^*(x) i \overleftrightarrow{\partial_0} \phi_{\text{KG}}(x) \, d^3 x$

we have

$$a(\omega_{\mathbf{k},\xi}, \mathbf{k}) = \frac{1}{2(\pi\omega_{\mathbf{k},\xi})^{1/2}} \int_0^\infty d\xi' \int d^4 x \, g_{\mathbf{k},\xi}^*(x) i \overleftrightarrow{\partial_0} \phi(x, \xi')$$

$$= \left(\frac{\omega_{\mathbf{k},\xi}}{\pi}\right)^{1/2} \int_0^\infty d\xi' \int d^4 x \, g_{\mathbf{k},\xi}^*(x) \phi(x, \xi') \tag{45}$$

Similarly, for antiparticles,

$$b(\omega_{\mathbf{k}}, \mathbf{k}) = \frac{1}{2\sqrt{\pi\omega_{\mathbf{k},\xi}}} \int_0^\infty d\xi' \int d^4x\, g^*_{\mathbf{k},\xi}(x) i \overleftrightarrow{\partial_0} \phi^\dagger(x,\xi'). \tag{46}$$

These expressions for the ladder operators are in addition to those provided above, Eqs. (35) and (36).

The Hamiltonian (33) involves an integration over the mass variable $\xi$ and is different than the usual one. Already at the classical level, we may ask what the implications are. With the expressions $H = \Pi\dot{\phi} + \Pi^*\dot{\phi}^* - L$ and $\Pi = \frac{\partial L}{\partial \dot{\phi}}$, the Lagrangian is found to be

$$L = \dot{\phi}^*\dot{\phi} - \nabla\phi^* \cdot \nabla\phi - \xi\phi^*\phi. \tag{47}$$

A variational analysis with fixed boundary, which follows the conventional treatment [17], can be undertaken to find the Euler-Lagrange equations that result from $L$. For this purpose, the action is defined as

$$S[\phi] = \int_R L\, d^4x\, d\xi, \tag{48}$$

where the integral is performed over a five-dimensional region $R$ having boundary $\partial R$ characterized by the equation $B(x,\xi) = 0$. We can consider a change in the fields $\phi \to \phi + \delta\phi(x,\xi)$ and $\phi^* \to \phi^* + \delta\phi^*(x,\xi)$, where $\delta\phi(x,\xi)$ and $\delta\phi^*(x,\xi)$ vanish on $\partial R$. A four-dimensional region $V$ has a boundary $\partial V$ formed by the intersection of the four-dimensional hyperplane $\xi = \xi'$ and $\partial R$, which boundary $\partial V$ is a three-dimensional hypersurface characterized by $B(x,\xi') = 0$. By construction, $\delta\phi(x,\xi)$ and $\delta\phi^*(x,\xi)$ also vanish on $\partial V$. The changes in the fields elicit a corresponding change $\delta S$ in the action, and the variational principle we take is

$$\delta S = \int_R \delta L\, d^4x\, d\xi = 0 \tag{49}$$

where

$$\delta L = \frac{\partial L}{\partial \phi}\delta\phi + \frac{\partial L}{\partial(\partial_\mu\phi)}\delta(\partial_\mu\phi) + \frac{\partial L}{\partial \phi^*}\delta\phi^* + \frac{\partial L}{\partial(\partial_\mu\phi^*)}\delta(\partial_\mu\phi^*). \tag{50}$$

Whence,

$$\begin{aligned}\int_R \left[\left(\frac{\partial L}{\partial \phi} - \partial_\mu \frac{\partial L}{\partial(\partial_\mu\phi)}\right)\delta\phi + \left(\frac{\partial L}{\partial \phi^*} - \partial_\mu \frac{\partial L}{\partial(\partial_\mu\phi^*)}\right)\delta\phi^*\right] d^4x\, d\xi \\ + \int_0^\infty \int_{\partial V} \left(\frac{\partial L}{\partial(\partial_\mu\phi)}\delta\phi + \frac{\partial L}{\partial(\partial_\mu\phi^*)}\delta\phi^*\right) d\sigma_\mu\, d\xi = 0\end{aligned} \tag{51}$$

where the four-dimensional divergence theorem [17] has been used on the last term, which can then be seen to vanish because $\delta\phi = \delta\phi^* = 0$ on the boundary $\partial V$.

Since $\delta\phi$ and $\delta\phi^*$ are arbitrary, the condition for Eq. (49) to hold is

$$\frac{\partial L}{\partial \phi} - \left(\partial_\mu \frac{\partial L}{\partial(\partial_\mu\phi)}\right) = 0 \tag{52}$$

and

$$\frac{\partial L}{\partial \phi^*} - \partial_\mu \left(\frac{\partial L}{\partial(\partial_\mu\phi^*)}\right) = 0, \tag{53}$$

resulting in the usual Klein-Gordon equations

$$(\partial^\mu\partial_\mu + \xi)\phi = 0 \tag{54}$$

and

$$(\partial^\mu\partial_\mu + \xi)\phi^* = 0. \tag{55}$$

### III. Mass-Independent Klein-Gordon Equation

There is a link between the Klein-Gordon equation and Laplace transforms that can be exploited when $\xi$ is variable. As a preliminary remark, if $\psi(x,\xi)$ is a solution of the Klein-Gordon equation for any $\xi > 0$, then the convolution

$$[\psi(x) * \psi^\dagger(x)](\xi) \equiv \int_0^\xi \psi(x, \xi - \xi')\psi^\dagger(x, \xi') \, d\xi' \tag{56}$$

is a solution of the Klein-Gordon equation with added source $2(\partial_\mu \psi * \partial^\mu \psi^\dagger)(\xi)$. Choosing the particular operator $\phi$ of Eq. (34) for $\psi$, the convolution theorem yields

$$\left(\partial^\mu \partial_\mu - \frac{d}{d\tau}\right)[\Phi(x,\tau)\Phi^\dagger(x,\tau)] = 2\partial^\mu \Phi(x,\tau)\partial_\mu \Phi^\dagger(x,\tau) \tag{57}$$

where the Laplace transform has been introduced,

$$\begin{aligned}\Phi(x,\tau) &= \mathcal{L}_\tau[\phi(x,\xi)] \\ &= \int_0^\infty e^{-\tau\xi}\phi(x,\xi)\,d\xi\end{aligned} \tag{58}$$

with $\tau \geq 0$.

Continuing in this vein, if the Klein-Gordon equation

$$-\Box\phi(x,\xi) = \xi\phi(x,\xi), \tag{59}$$

with $\Box = \partial^\mu \partial_\mu$ and $\xi = \frac{m}{\hbar^2}E_0$, is viewed as a (second quantized) relativistic analog of the time independent Schrodinger equation,

$$-\Delta\psi\left(x,\tilde{\xi}\right) = \tilde{\xi}\psi\left(x,\tilde{\xi}\right), \tag{60}$$

with $\Delta = -\partial^j \partial_j$ and $\tilde{\xi} = \frac{2m}{\hbar^2}E$, then two further relations are suggested.

First, in analogy with how a Fourier transform (supplemented with initial/boundary conditions) converts Eq. (60) to the time-dependent Schrodinger equation

$$\frac{1}{i}\frac{d}{d\tilde{\tau}}\Psi(\tilde{\tau},x) = \Delta\Psi(\tilde{\tau},x) \tag{61}$$

where

$$\Psi(\tilde{\tau},x) = \int e^{-i\tilde{\xi}\tilde{\tau}}\psi\left(x,\tilde{\xi}\right)d\tilde{\xi} \tag{62}$$

and $\tilde{\tau} = \frac{\hbar}{2m} t$,

a Laplace transform converts Eq. (59) to the mass-independent Klein-Gordon equation (MIKE)

$$\frac{d}{d\tau} \Phi(x, \tau) = \Box \Phi(x, \tau). \tag{63}$$

Comparing the time-dependent Schrodinger equation to MIKE, we see that $\frac{1}{i}\frac{d}{d\tilde{\tau}}$ has been replaced by $\frac{d}{d\tau}$, and the Laplacian $\Delta$ has been replaced by the Minkowski Laplacian $\Box$, which is conducive to a covariant formulation where time and space are treated more symmetrically. (Feynman [6] remarked that Eq. (6) is analogous to the time-dependent Schrodinger equation.) The absence of $i$ in MIKE stems from the fact that we Laplace instead of Fourier transformed to keep $\xi = m^2$ positive, thereby avoiding imaginary masses and virtual particles arising from spacelike four-momenta. (Eq. (1), in contrast to MIKE, does include a factor $i$, a result of involving the Fourier, instead of Laplace, transform---more on this later.) The Lorentz invariant parameter $\tau$, conjugate to $\xi$, plays the role of a time, and the "initial" condition for MIKE is

$$\Phi(x, 0) = \int_0^\infty \phi(x, \xi) \, d\xi \tag{64}$$

where $\phi(x, \xi)$ is the solution of the Klein-Gordon equation (8), which, as mentioned above, is somewhat different than $\phi_{KG}$ due to the different ladder operators. The field operator $\Phi^\dagger(x, 0)$ creates a single particle state according to (cf. Eq. 5.18 in [5])

$$(2\pi)^{3/2} \Phi^\dagger(x, 0)|0\rangle = |x\rangle. \tag{65}$$

We can further exploit these analogies by writing MIKE in operator form

$$-\frac{d}{d\tau}|\widehat{\Phi}(\tau)\rangle = \hat{k}^\mu \hat{k}_\mu |\widehat{\Phi}(\tau)\rangle \tag{66}$$

where

$$|\widehat{\Phi}(\tau)\rangle = \exp(-\tau \hat{k}^\mu \hat{k}_\mu)|\widehat{\Phi}(0)\rangle \tag{67}$$

with

$$\langle x|\hat{k}^\mu \hat{k}_\mu|\widehat{\Phi}(\tau)\rangle = -\Box \Phi(x,\tau) \tag{68}$$

and $\Phi(x,\tau) = \langle x|\widehat{\Phi}(\tau)\rangle$. (An explicit expression for $|\widehat{\Phi}(\tau)\rangle$ is provided in Eq. (76) below.) The hybrid notation $|\widehat{\Phi}(\tau)\rangle$ comes from single-particle relativistic quantum mechanics: if $\langle x|\phi_{KG}^{(+)}\rangle = \frac{1}{(2\pi)^3}\int \langle x|k\rangle \Phi_{KG}^{(+)}(k)\delta(k^2 - m^2)\theta(k^0)\, d^4k$ is a solution of the Klein-Gordon equation written using kets, then after second quantization, wherein we replace the complex-valued function $\Phi_{KG}^{(+)}(k)$ by the operator $a_{KG}(k)$, we can write $\langle x|\hat{\phi}_{KG}^{(+)}\rangle = \frac{1}{(2\pi)^3}\int \langle x|k\rangle a_{KG}(k)\delta(k^2 - m^2)\theta(k^0)\, d^4k$. (In keeping with standard notation, we have usually omitted carets on ladder operators $a, b$ and field operators $\phi$ in this paper, though we include them when fields appear in kets.) Then,

$$|\widehat{\Phi}(\tau)\rangle = \int_0^\infty e^{-\xi\tau}|\hat{\phi}(\xi)\rangle d\xi \tag{69}$$

where

$$|\hat{\phi}(\xi)\rangle = \frac{1}{(2\pi)^{3/2}}\int |k\rangle a(k)\delta(k^2 - \xi)\theta(k^0)d^4k + \frac{1}{(2\pi)^{3/2}}\int |-k\rangle b^\dagger(k)\delta(k^2 - \xi)\theta(k^0)d^4k \tag{70}$$

and

$\langle x|\hat{\phi}(\xi)\rangle = \phi(x,\xi)$. Introducing the "bra" $\langle \widehat{\Phi}(\tau)|$, satisfying $\langle \widehat{\Phi}(\tau)|x\rangle = [\widehat{\Phi}(x,\tau)]^\dagger$, allows us to write

$$\frac{d}{d\tau}\hat{\rho}(\tau) = -\left[\hat{k}^\mu \hat{k}_\mu, \hat{\rho}(\tau)\right]_+ \tag{71}$$

where

$$\hat{\rho}(\tau) = |\hat{\Phi}(\tau)\rangle\langle\hat{\Phi}(\tau)|. \tag{72}$$

Eq. (71) is reminiscent of the von Neumann equation from nonrelativistic quantum mechanics, although here it is the anticommutator that enters (note the "+" subscript).

Second, in analogy with the quantum Liouville eigenvalue equation $\left(\frac{1}{\hbar}[\hat{\mathbf{p}} \cdot \hat{\mathbf{p}},\cdot] - \underline{\lambda}\right)\underline{\hat{\sigma}}(\lambda) = 0$, we can write

$$\left([\hat{p}^\mu \hat{p}_\mu,\cdot]_+ - \lambda\right)\hat{\sigma}(\lambda) = 0. \tag{73}$$

The $\lambda$ and the $\hat{\sigma}$ in this last equation are the eigenvalues and field operator eigensolutions of the operator $[\hat{p}^\mu \hat{p}_\mu,\cdot]_+$. In the position representation,

$$(-\Box_x - \Box_y - \lambda)\hat{\sigma}(x,y;\lambda) = 0. \tag{74}$$

We can confirm that

$$(-\Box_x - \Box_y - \xi)[\hat{\phi}(x) * \hat{\phi}^\dagger(y)](\xi) = 0 \tag{75}$$

where $\hat{\phi}$ is a solution of the Klein-Gordon equation. That is, the convolutions $(|\hat{\phi}\rangle * \langle\hat{\phi}|)(\xi)$ are field operator eigensolutions of the operator $[\hat{p}^\mu \hat{p}_\mu,\cdot]_+$ with eigenvalues $\xi > 0$.

When viewed as a single particle wave function, we can solve MIKE by a separation of variables, similar to the procedure for the Klein-Gordon equation; however, five separation constants arise instead of the conventional four, namely the four momenta $k^\mu$ and a fifth constant constrained to equal $-k^2$.

Having "freed up" the time variable in MIKE, we permit offshell values of $k^\mu$, but with two restrictions. We take positive energy solutions and causal $k$, and correspondingly insert $\theta(k^0)$ and $\theta(k^2)$ in the following second quantized expression

$$|\hat{\Phi}(\tau)\rangle = \frac{1}{(2\pi)^{3/2}} \int |k\rangle a(k) e^{-k^2\tau} \theta(k^0)\theta(k^2) d^4k$$
$$+ \frac{1}{(2\pi)^{3/2}} \int |-k\rangle b^\dagger(k) e^{-k^2\tau} \theta(k^0)\theta(k^2) d^4k, \quad (76)$$

which implies

$$\langle\hat{\Phi}(\tau)| = \frac{1}{(2\pi)^{\frac{3}{2}}} \int a^\dagger(k)\langle k| e^{-k^2\tau} \theta(k^0)\theta(k^2) d^4k$$
$$+ \frac{1}{(2\pi)^{\frac{3}{2}}} \int b(k)\langle -k| e^{-k^2\tau} \theta(k^0)\theta(k^2) d^4k. \quad (77)$$

The values $k^2$ and $\tau$ are analogous to the frequency $\omega_{\mathbf{k}}$ and the time $x^0$, respectively, except that the former pair are invariant, and the latter pair are not. For later, we note that in Eqs. (76) and (77), the $k^2$ argument of the step function $\theta(k^2)$ corresponds to one-half the diagonal eigenvalues of the operator relation,

$$[\hat{k}^\mu \hat{k}_\mu, |k\rangle\langle k'|]_+ = (k^2 + k'^2)|k\rangle\langle k'|. \quad (78)$$

Using the identity $\mathcal{L}_\xi^{-1}\left[e^{-k^2\tau}\theta(k^2)\right] = \delta(\xi - k^2)$, we may take the inverse Laplace transform of Eq. (76) to recover Eq. (70).

Alternatively, in the position representation, Eq. (76) becomes

$$\Phi(x,\tau) = \frac{1}{(2\pi)^{3/2}} \int a(k) e^{-k^2\tau} e^{-ikx} \theta(k^0)\theta(k^2) d^4k$$
$$+ \frac{1}{(2\pi)^{3/2}} \int b^\dagger(k) e^{-k^2\tau} e^{ikx} \theta(k^0)\theta(k^2) d^4k, \quad (79)$$

the inverse Laplace transform of which, when evaluated at $\xi = \xi_p$, yields Eq. (24).

We note that because $\Phi(x,\tau)$ is independent of $\xi$, so too are the operators $a(k)$ and $b(k)$, which therefore must be different from $a_{KG}(\mathbf{k})$ and $b_{KG}(\mathbf{k})$ which do depend on $\xi$. Consistent with $a(k)|0\rangle = 0$, $a^\dagger(k)|0\rangle = |k\rangle$ and $\langle x|k\rangle = e^{-ikx}$, we have $\langle k'|k\rangle = \langle 0|[a(k'), a^\dagger(k)]|0\rangle = \langle 0|[b(k'), b^\dagger(k)]|0\rangle = \delta^4(k' - k)$.

An advantage of using a hybrid field, involving $a(k)$ and $b(k)$ with the $\delta(k^2 - \xi)\theta(k^0)$ constraint, is that some kernels arising from pairs of field operators (two-point correlators) are manifestly positive in both $\xi$ and $\tau$ space. The notion of a positive kernel in the context of two-point correlators has been discussed previously [18]. Briefly, the complex valued kernel $\sigma(x, y)$ of two spacetime variables may be used to define a functional $\sigma[f]$, where $f(x)$ is a complex function, according to

$$\sigma[f] = \int\int f^*(x)\sigma(x,y)f(y)\, d^4x d^4y \qquad (80)$$

The functional is positive semidefinite ("positive" for short, when there is no ambiguity) if $\sigma[f] \geq 0$ for all integration regions and test functions $f$ for which the integral exists, and we succinctly write $\sigma \geq 0$.

If

$$\operatorname{Re}\int\int f^*(x)\sigma(x,y)f(y)\, d^4x d^4y \geq 0 \qquad (81)$$

or

$$\operatorname{Im}\int\int f^*(x)\sigma(x,y)f(y)\, d^4x d^4y \geq 0, \qquad (82)$$

we write $\operatorname{Re}\sigma \geq 0$ or $\operatorname{Im}\sigma \geq 0$, respectively. These notions can be traced back at least as far as Hilbert [19].

A kernel is manifestly positive if we can write it is as

$$\sigma(x,y) = \int du\, p(u) \langle \chi | \hat{A}(x,u) \hat{A}^\dagger(y,u) | \chi \rangle \tag{83}$$

where $p(u) \geq 0$, and $\hat{A}$ is an operator and $|\chi\rangle$ a ket (such as the vacuum) in Fock space.

An example of physical import is $\operatorname{Re} i\Delta_F \geq 0$ [18], where the well-known Feynman propagator $\Delta_F(x,y)$ is given by

$$i\Delta_F(x,y) = \langle 0 | T\{\phi_{KG}(x)\phi^\dagger_{KG}(y)\} | 0 \rangle$$

$$= \begin{cases} \int e^{-i\omega_k(x^0-y^0)} e^{i\mathbf{k}\cdot(\mathbf{x}-\mathbf{y})} \dfrac{d^3k}{(2\pi)^3 2\omega_\mathbf{k}} & \text{if } y^0 < x^0 \\ \int e^{i\omega_k(x^0-y^0)} e^{-i\mathbf{k}\cdot(\mathbf{x}-\mathbf{y})} \dfrac{d^3k}{(2\pi)^3 2\omega_\mathbf{k}} & \text{if } x^0 < y^0, \end{cases} \tag{84}$$

$T$ being the time ordering operator. The Feynman propagator has been examined using parametric time in [3] and [20].

Let us examine what we mean by saying that a kernel is manifestly positive in both $\xi$ and $\tau$ space by considering the manifestly positive kernel $\langle 0 | \phi_{KG}(x,\xi) \phi^\dagger_{KG}(y,\xi) | 0 \rangle$. A computation reveals that we may express this kernel as a convolution in the new field operators according to

$$\langle 0 | \phi_{KG}(x) \phi^\dagger_{KG}(y) | 0 \rangle = 2 \langle 0 | [\phi(x) * \phi^\dagger(y)](2\xi) | 0 \rangle. \tag{85}$$

An application of the convolution theorem then implies

$$\mathcal{L}_{2\tau} \langle 0 | \phi_{KG}(x,\xi) \phi^\dagger_{KG}(y,\xi) | 0 \rangle = \langle 0 | \Phi(x,\tau) \Phi^\dagger(y,\tau) | 0 \rangle. \tag{86}$$

The left-hand side is a manifestly positive kernel, and this manifest property is preserved on the right-hand side.

Although the Laplace transform of a positive kernel is positive, this is not necessarily true for the inverse Laplace transform. What we have then is a necessary but not sufficient condition for positivity: we must draw from the set of positive kernels in $\tau$ for the associated kernels in $\xi$---obtained by inverse Laplace transforming---to also be positive. This partly explains why completely positive maps in the variable $\tau$, which we will consider below, are attractive.

Several other expressions involving pairs of conventional field operators may be related to convolutions in the new field operators. For example, by direct computation we can verify that

$$\left[\phi_{\text{KG}}(t,\mathbf{x}), \dot{\phi}^\dagger_{\text{KG}}(t,\mathbf{y})\right] = 2\int_0^{2\xi}\left[\phi(t,\mathbf{x}; 2\xi - \xi'), \dot{\phi}^\dagger(t,\mathbf{y};\xi')\right]d\xi' \qquad (87)$$

Again using the convolution theorem, we obtain

$$-i\mathcal{L}_{2\tau}\left[\phi_{\text{KG}}(t,\mathbf{x};\xi), \dot{\phi}^\dagger_{\text{KG}}(t,\mathbf{y};\xi)\right] = -i\left[\Phi(t,\mathbf{x};\tau), \dot{\Phi}^\dagger(t,\mathbf{y};\tau)\right] \qquad (88)$$

The left-hand side is a positive kernel on account of

$$\left[\phi_{\text{KG}}(t,\mathbf{x};\xi), \dot{\phi}^\dagger_{\text{KG}}(t,\mathbf{y};\xi)\right] = i\delta(\mathbf{x}-\mathbf{y}), \qquad (89)$$

and hence so too is the right-hand side.

Another example, which we will examine further in the following sections, is the Feynman propagator of Eq. (84), which we can write as a convolution

$$\begin{aligned} i\Delta_{\text{F}}(x,y) &= 2\langle 0|T[\phi(x)*\phi^\dagger(y)](2\xi)|0\rangle \\ &= 2\mathcal{L}^{-1}_{2\xi}\langle 0|T\langle x|\rho(\tau)|y\rangle|0\rangle. \end{aligned} \qquad (90)$$

## IV. Interacting Systems

We now turn to scalar electrodynamics, an interacting system in which a spinless particle is coupled to the electromagnetic field. With the usual prescription [11] of replacing the derivative $\partial$ with the covariant derivative $D = \partial + ieA$ in the Klein-Gordon equations (8) and (9), we obtain the coupled equations of scalar electrodynamics,

$$[(iD)^2 - m^2]\phi = 0 \tag{91}$$

$$[(iD)^{*2} - m^2]\phi^\dagger = 0 \tag{92}$$

and

$$\Box A^\mu = eJ^\mu \tag{93}$$

where

$$J^\mu = \frac{1}{2}\left(\phi^\dagger(iD^\mu\phi) + (iD^\mu\phi)\phi^\dagger\right) + \frac{1}{2}\left(\phi(iD^\mu\phi)^\dagger + (iD^\mu\phi)^\dagger\phi\right). \tag{94}$$

Laplace transforming with respect to $\xi$,

$$\frac{d}{d\tau}\Phi(\tau) = -\mathcal{L}_\tau\left[(i\partial - eA)^2 \mathcal{L}_\xi^{-1}\Phi(\tau)\right] \tag{95}$$

or in ket form

$$\frac{d}{d\tau}|\widehat{\Phi}(\tau)\rangle = -\mathcal{L}_\tau\left[\left(\hat{p} - e\hat{A}(\hat{x})\right)^2 \mathcal{L}_\xi^{-1}|\widehat{\Phi}(\tau)\rangle\right] \tag{96}$$

where

$$\Phi(\tau) = \langle x|\widehat{\Phi}(\tau)\rangle$$
$$= \int_0^\infty d\xi\, e^{-\xi\tau} \int \frac{d^3k}{(2\pi)^{3/2} 2\omega_{\mathbf{k},\xi}} \times \tag{97}$$

$$\left[U_I^\dagger(t)a_S(\omega_{\mathbf{k},\xi},\mathbf{k})U_I(t)\langle x|\omega_{\mathbf{k},\xi},\mathbf{k}\rangle + U_I^\dagger(t)b_S^\dagger(\omega_{\mathbf{k},\xi},\mathbf{k})U_I(t)\langle x|-\omega_{\mathbf{k},\xi},-\mathbf{k}\rangle\right]$$

with "S" denoting the Schrodinger picture and $U_I(t)$ the interaction picture propagator. Eq. (95) or Eq. (96) is MIKE for scalar electrodynamics.

In view of Eq. (90), we define a noisy Feynman propagator according to

$$i\Delta_{F,n}(x,y) = 2\mathcal{L}_{2\xi}^{-1}\text{Tr}_R\sigma_R\langle 0|T\langle x|\rho(\tau)|y\rangle|0\rangle \tag{98}$$

where $\langle x|\rho(\tau)|y\rangle$ is obtained from Eq. (72), except that here Heisenberg operators evolve under the coupled Hamiltonian of scalar electrodynamics; the trace is over electromagnetic variables and $\sigma_R$ is an appropriate "reservoir," or an initial electromagnetic density operator.

Because the equations (91)-(93) are coupled, the electromagnetic field $\hat{A}(x)$ can depend on parameters of the Klein-Gordon particle. One helpful approximation that we may invoke is to ignore the functional dependence of $\hat{A}(x)$ on the mass of the Klein-Gordon particle to which the electromagnetic field is coupled. If that were done, MIKE for scalar electrodynamics takes on the simpler form

$$\frac{d}{d\tau}|\widehat{\Phi}(\tau)\rangle = -\left(\hat{p} - e\hat{A}(\hat{x})\right)^2 |\widehat{\Phi}(\tau)\rangle, \tag{99}$$

which may be compared to Eq. (6). The formal solution of Eq. (99) is

$$|\widehat{\Phi}(\tau)\rangle = \exp\left[-\tau\left(\hat{p} - e\hat{A}(\hat{x})\right)^2\right] |\widehat{\Phi}(0)\rangle \tag{100}$$

with initial condition $|\widehat{\Phi}(0)\rangle$ given by

$$\begin{aligned}\langle x|\widehat{\Phi}(0)\rangle &= \int_0^\infty \langle x|\phi(\xi)\rangle\, d\xi \\ &= \int_0^\infty d\xi \int \frac{d^4k}{(2\pi)^{3/2}} \left[a_H(\mathbf{k},t)\langle x|0,\mathbf{k}\rangle + b_H^\dagger(\mathbf{k},t)\langle x|0,-\mathbf{k}\rangle\right]\delta(k^2-\xi)\theta(k^0) \\ &= \int_0^\infty d\xi \int \frac{d^3k}{(2\pi)^{3/2}2\omega_{\mathbf{k},\xi}} \begin{bmatrix} U_I^\dagger(t)a_S(\omega_{\mathbf{k},\xi},\mathbf{k})U_I(t)\langle x|\omega_{\mathbf{k},\xi},\mathbf{k}\rangle \\ +U_I^\dagger(t)b_S^\dagger(\omega_{\mathbf{k},\xi},\mathbf{k})U_I(t)\langle x|-\omega_{\mathbf{k},\xi},-\mathbf{k}\rangle\end{bmatrix}\end{aligned} \tag{101}$$

The quantity $\Phi(x,0)$ is not much easier to compute than $\Phi(x,\tau)$ because both require knowledge of the field operator $\phi(x,\xi)$ in the Heisenberg picture. Later on, we will consider a perturbative solution of $|\widehat{\Phi}(0)\rangle$, in particular to zeroth order.

The approximate MIKE for scalar electrodynamics (Eq. (99)) gives rise to a particular form for the noisy Feynman propagator that, before inverse Laplace transforming, is reminiscent of completely positive evolution. (We recall that in non-relativistic quantum mechanics, completely positive evolution of the density operator is via a map $\sum_\alpha X_\alpha^\dagger(t) \cdot X_\alpha(t)$ [16], [21].) To see this, we can utilize the assumed positivity of $\sigma_R$ to take its square root:

$$i\Delta_{F,n}(x,y) = 2\mathcal{L}_{2\xi}^{-1}\text{Tr}_T\left[\left(\sigma_R^{1/2}\otimes|0\rangle\langle0|\right)T\{\langle x|\rho(\tau)|y\rangle\}\left(\sigma_R^{1/2}\otimes 1_{KG}\right)\right] \quad (102)$$

where

$$\langle x|\rho(\tau)|y\rangle = \langle x|e^{-\tau\left(\hat{p}-e\hat{A}(\hat{x})\right)^2}\rho(0)e^{-\tau\left(\hat{p}-e\hat{A}(\hat{x})\right)^2}|y\rangle, \quad (103)$$

$$\rho(0) = |\widehat{\Phi}(0)\rangle\langle\widehat{\Phi}(0)| \quad (104)$$

and the subscript T indicates a trace over the total complement of electromagnetic and Klen-Gordon variables. To trace over the electromagnetic part, we can use an occupation number basis in Fock space, which we can schematically write as $|\{n_i\}\rangle$. The occupation numbers extend over the transverse and longitudinal/scalar number states [22].

If we assume that Hermitian $\sigma_R$ is diagonal in this basis with

$$\sigma_R|\{n_j\}\rangle = \lambda_{\{n_j\}}|\{n_j\}\rangle, \quad (105)$$

$$\langle\{m_i\}|\{n_i\}\rangle = \delta_{\{m_i\},\{n_i\}} \quad (106)$$

and

$$\sum_{\{n_i\}}|\{n_i\}\rangle\langle\{n_i\}| = 1_R, \quad (107)$$

we can write

$$i\Delta_{F,n}(x,y) = 2\mathcal{L}_{2\xi}^{-1}\sum_{\{n_i\}}\langle 0|\langle\{n_i\}|\lambda_{\{n_i\}}^{1/2}T\{\langle x|\rho(\tau)|y\rangle\}\lambda_{\{n_i\}}^{1/2}|\{n_i\}\rangle|0\rangle \quad (108)$$

Eq. (108) can be made to resemble completely positive evolution more closely if we make a further approximation. The field operator $\hat{A}(x) = U_I^\dagger(t)\hat{A}_I(x)U_I(t)$ and the ladder operator $a(x) = U_I^\dagger(t)a_I(x)U_I(t)$ appearing in $\langle x|\rho(\tau)|y\rangle$ (see Eqs. (101) and (103)) are in the Heisenberg picture; the simplest approximation we can subsequently make is a zeroth approximation for $U_I(t)$ wherein the interaction picture operator is replaced by the unit operator. In such case,

$$(|\hat{\Phi}(0)\rangle\langle\hat{\Phi}(0)|)(|\{n_i\}\rangle\langle\{n_i\}|\otimes 1_{KG}) = |\{n_i\}\rangle|\hat{\Phi}(0)\rangle\langle\hat{\Phi}(0)|\langle\{n_i\}| \qquad (109)$$

since with this approximation the ladder operators in $|\Phi(0)\rangle$ evolve freely according to $a_I(x) = e^{-ik^0 x^0} a$ and $b_I^\dagger(x) = e^{ik^0 x^0} b^\dagger$. We would then obtain

$$i\Delta_{F,n}^{(0)}(x,y)$$
$$= 2\mathcal{L}_{2\xi}^{-1} \sum_{\{m_i\},\{n_i\}} \langle 0|T\langle x|\hat{W}_{\{m_i\},\{n_i\}}^\dagger(\tau)|\hat{\Phi}(0)\rangle\langle\hat{\Phi}(0)|\hat{W}_{\{m_i\},\{n_i\}}(\tau)|y\rangle|0\rangle \qquad (110)$$

where

$$\hat{W}_{\{m_i\},\{n_i\}}(\tau) = \langle\{m_i\}|\exp\left[-\tau\left(\hat{p} - e\hat{A}_I(\hat{x})\right)^2\right]\lambda_{\{n_i\}}^{1/2}|\{n_i\}\rangle \qquad (111)$$

and

$$\hat{W}_{\{m_i\},\{n_i\}}^\dagger(\tau) = \langle\{n_i\}|\lambda_{\{n_i\}}^{1/2}\exp\left[-\tau\left(\hat{p} - e\hat{A}_I(\hat{x})\right)^2\right]|\{m_i\}\rangle. \qquad (112)$$

Whence,

$$\langle x|\hat{W}_{\{m_i\},\{n_i\}}^\dagger(\tau)|\hat{\Phi}(0)\rangle$$
$$= \langle\{n_i\}|\lambda_{\{n_i\}}^{1/2}\exp\left[-\tau\left(i\partial_x - e\hat{A}_I(x)\right)^2\right]|\{m_i\}\rangle\Phi(x,0) \qquad (113)$$

and

$$\langle\hat{\Phi}(0)|\hat{W}_{\{m_i\},\{n_i\}}(\tau)|y\rangle$$
$$= \left\{\langle\{n_i\}|\lambda_{\{n_i\}}^{1/2}\exp\left[-\tau\left(i\partial_y - e\hat{A}_I(y)\right)^2\right]|\{m_i\}\rangle\Phi(y,0)\right\}^\dagger. \qquad (114)$$

Some aspects of the noisy Feynman propagator will be investigated in the next section.

## V. Sign of the Noisy Feynman Propagator

The Feynman propagator possesses the important positivity property $\operatorname{Re} i \Delta_F \geq 0$ (see Section III). To investigate whether this property continues to hold in a model having a dynamical map similar in form to the noisy Feynman propagator of Eq. (110), we consider

$$\operatorname{Re} i \overline{\Delta}_{F,n}^{(0)}(x,y) \equiv 2\operatorname{Re}\mathcal{L}_{2\xi}^{-1} \sum_{l} \langle 0|T\langle x|\hat{V}_l^\dagger(\tau)|\hat{\Phi}(0)\rangle\langle\hat{\Phi}(0)|\hat{V}_l(\tau)|y\rangle|0\rangle. \tag{115}$$

We recall that Eq. (110) was obtained in a zeroth order approximation of $U_I(t)$ in which the operator $\hat{W}_{\{m_i\},\{n_i\}}(\tau)$ is a function of $\hat{x}$ and $\hat{k}$, but no longer a function of second quantized operators. Likewise, we shall assume the same for the operator $\hat{V}_l(\tau)$. The overbar in $\overline{\Delta}_{F,n}^{(0)}(x,y)$ indicates that in the phenomenological expression that is Eq. (115), $\hat{W}_{\{m_i\},\{n_i\}}(\tau)$ has been replaced by $\hat{V}_l(\tau)$. In Equation (115),

$$|\hat{\Phi}(0)\rangle = \frac{1}{(2\pi)^{3/2}} \int |k\rangle a(k)\, \theta(\lambda(k)/2)\theta(k^0) d^4k$$
$$+ \frac{1}{(2\pi)^{3/2}} \int |-k\rangle b^\dagger(k)\, \theta(\lambda(k)/2)\theta(k^0) d^4k \tag{116}$$

where the ladder operators are in the Schrodinger picture. Later, we will identify $\lambda(k)$ with the diagonal elements of a Liouville operator. We note that for the free field case, $\lambda(k) = 2k^2$, leading to the $\theta(k^2)$ factor in Eq. (76).

Continuing, we may then compute
$$\langle x|\hat{V}_l^\dagger(\tau)|\hat{\Phi}(0)\rangle$$
$$= \frac{1}{(2\pi)^{3/2}} \int a(k)\overline{V}_l^*(k,x,\tau)\theta(\lambda(k)/2)\theta(k^0) d^4k$$
$$+ \frac{1}{(2\pi)^{3/2}} \int b^\dagger(k)\overline{V}_l^*(-k,x,\tau)\theta(\lambda(k)/2)\theta(k^0) d^4k \tag{117}$$

where $\bar{V}_l(k, x, \tau) = \langle k|\hat{V}_l(\tau)|x\rangle$.

Whence,

$$\text{Re}i\bar{\Delta}_{F,n}^{(0)}(x,y) = \frac{2}{(2\pi)^3}\sum_l \text{Re}\mathcal{L}_{2\xi}^{-1}\int d^4k\ \theta(\lambda(k)/2)\ \theta(k^0) \times \tag{118}$$

$$[\theta(x^0-y^0)\bar{V}_l^*(k,x,\tau)\bar{V}_l(k,y,\tau) + \theta(y^0-x^0)\bar{V}_l^*(-k,x,\tau)\bar{V}_l(-k,y,\tau)]$$

If the summand is a positive kernel, so too is $\text{Re}\,i\,\bar{\Delta}_{F,n}^{(0)}(x,y)$.

Introducing the real four-vector $\zeta$, we may consider an example in which

$$\hat{V}(\tau) = e^{-\tau\hat{k}^2}e^{-u\zeta\hat{k}\tau^{1/2}}, \tag{119}$$

$$\lambda(k) = 2[k^2 - 2(\zeta k)^2] \tag{120}$$

and $\sum_{l=1}^{N} \to \frac{1}{2\sqrt{\pi}}\int_{-\infty}^{\infty}du\,e^{-u^2/4}$. (Recall that in the notation we are using, $\zeta k \equiv \zeta^\mu k_\mu$.) We assume that $1 - 2(\zeta^0)^2 > 0$. Let us pause to somewhat motivate these choices. The evolution equation

$$\sigma(\tau) = \frac{1}{2\sqrt{\pi}}\int_{-\infty}^{\infty} du\,e^{-u^2/4}e^{-u\zeta\hat{k}\tau^{1/2}}\sigma(0)e^{-u\zeta\hat{k}\tau^{1/2}} \tag{121}$$

implies

$$\frac{d\sigma}{d\tau} = \left[\zeta\hat{k},[\zeta\hat{k},\sigma(\tau)]_+\right]_+ \tag{122}$$

suggestive of a theory that couples the momentum of a Klein-Gordon particle to another field $A$, such as a linear coupling $pA$.

Adding the free field portion (see Eq. (71)), we define the Liouville operator

$$L(\cdot) \equiv \left[\hat{k}^2,\cdot\right]_+ - \left[\zeta\hat{k},[\zeta\hat{k},\cdot]_+\right]_+ \tag{123}$$

with associated dynamical equation $d\rho/d\tau = -L(\rho)$. We note the following eigenvector equation

$$L(|k\rangle\langle k'|) = [k^2 + k'^2 - (\zeta k + \zeta k')^2]|k\rangle\langle k'| \qquad (124)$$

from which we see that the diagonal elements of $L$ are $2[k^2 - 2(\zeta k)^2]$. Restricting these elements to positive values gives rise to the argument of the step function in Eq. (118) with the choice for $\lambda(k)$ given by Eq. (120). We may then avail ourselves of the relation

$$\mathcal{L}_{2\xi}^{-1}\{e^{-2[k^2-2(\zeta k)^2]\tau}\theta[k^2 - 2(\zeta k)^2]\} = \frac{1}{2}\delta\{\xi - [k^2 - 2(\zeta k)^2]\} \qquad (125)$$

to calculate

$$i\overline{\Delta}_{F,n}^{(0)}(x,y) = \frac{1}{(2\pi)^3 2}\int \frac{d^3k}{\sqrt{(1-2(\zeta^0)^2)(\mathbf{k}^2+\xi)+2(\boldsymbol{\zeta}\cdot\mathbf{k})^2}} \times$$
$$[\theta(x^0 - y^0)e^{-i[\varpi(x^0-y^0)-\mathbf{k}\cdot(\mathbf{x}-\mathbf{y})]} + \theta(y^0 - x^0)e^{i[\varpi(x^0-y^0)-\mathbf{k}\cdot(\mathbf{x}-\mathbf{y})]}] \qquad (126)$$

where the positive frequency $\varpi$ is given by

$$\varpi(\mathbf{k}, \xi, \boldsymbol{\zeta}) =$$
$$\frac{1}{1 - 2(\zeta^0)^2}\left[\sqrt{(1 - 2(\zeta^0)^2)(\mathbf{k}^2 + \xi + 2(\boldsymbol{\zeta}\cdot\mathbf{k})^2) + (2\zeta^0\boldsymbol{\zeta}\cdot\mathbf{k})^2} - 2\zeta^0\boldsymbol{\zeta}\cdot\mathbf{k}\right]. \qquad (127)$$

To examine the positivity property of Re $i\overline{\Delta}_{F,n}^{(0)}$, we evaluate this functional at an arbitrary function $f$,

$$i\overline{\Delta}_{F,n}^{(0)}[f] = \frac{1}{(2\pi)^3 2}\int \frac{d^3k}{\sqrt{(1-2(\zeta^0)^2)(\mathbf{k}^2+\xi)+2(\boldsymbol{\zeta}\cdot\mathbf{k})^2}} \times \qquad (128)$$

$$\left\{ \begin{array}{l} \int d^4x \int d^4y\, f^*(x)f(y)[\cos(kx)\cos(ky) + \sin(kx)\sin(ky)] \\ -i\int d^4x \int d^4y f^*(x)f(y)[\theta(x^0-y^0) - \theta(y^0-x^0)]\sin[\varpi(x^0-y^0) - \mathbf{k}\cdot(\mathbf{x}-\mathbf{y})] \end{array} \right\}$$

By swapping the dummy variables $x$ and $y$, we can see that the second term in the brace brackets is purely imaginary, and hence

$$\operatorname{Re} i \overline{\Delta}_{F,n}^{(0)}[f] = \frac{1}{(2\pi)^3 2} \int \frac{d^3k}{\sqrt{(1-2(\zeta^0)^2)(\mathbf{k}^2+\xi)+2(\boldsymbol{\zeta}\cdot\mathbf{k})^2}} \times$$
$$\left[\left|\int f(x)\cos(kx)\,d^4x\right|^2 + \left|\int f(x)\sin(kx)\,d^4x\right|^2\right]. \tag{129}$$

We conclude that for this example, $\operatorname{Re} i \overline{\Delta}_{F,n}^{(0)} \geq 0$.

## VI. Discussion

Solutions $e^{-ikx}$ of the Klein-Gordon equation must satisfy the mass shell condition, usually expressed as the positive energy dispersion relation $E(m,\mathbf{k}) = \sqrt{\mathbf{k}^2 + m^2}$, which is reasonable since for an isolated elementary particle, which by definition has no internal structure, the mass is constant and logically appears on the right hand side of the last equation. However, for a massive composite particle, "[t]he mass of a body is not a constant; it varies with changes in its energy." [23] In such case, it is reasonable to cast the dispersion relation as $m(E,\mathbf{k},) = \sqrt{E^2 - \mathbf{k}^2}$ since the value of the Lorentz invariant $\xi = m^2$ can now change by varying both the composite's total momentum and energy. This conceptual difference between the two types of massive particles arises because an isolated elementary particle in a centre of mass reference frame has no kinetic energy, whereas the constituents of a composite particle may. An atom or a nucleus, for example, can exist in various excited states with different associated masses. The well-known relativistic result that the mass of a composite particle is not equal to the sum of the masses of its constituents is a reflection of this. It is not surprising then that variable mass theories in the literature have found favour in the study of anomalous low energy nuclear phenomena [4]. Once we isolate a system, however, its total energy cannot change; the relation $\xi = (k^0)^2$, valid in the centre of mass reference frame of a massive system, then implies that $\xi$ remains constant.

Three quantum fields included herein, involving $a_{KG}(\mathbf{k})\delta(k^2 - \xi_p)$, $a(k)\theta(k^2)$ and $a(k)\delta(k^2 - \xi_p)$, treat mass in somewhat different ways. In the conventional field $\phi_{KG}$ (Eq. (21)), raising operators create particles from the vacuum that are on-shell, restricting these particles to have a particular mass. On the other hand, the field $\phi_{PC}$ (Eq. (153)) can create particles with timelike 4-momenta, but with varying values of $k^2$ ($\langle \omega_{\mathbf{k},\xi}, \mathbf{k}|\phi^\dagger(x)|0\rangle$ is non-zero for any $\xi$). The field $\phi$ (Eq. (24)) is a hybrid, including creation operators associated with varying values of $\xi$, but also including a delta function that restricts $\xi$ to a particular value $\xi_p$. Expressions may be derived that interrelate these fields. Among other advantages, the theory presented here affords new ways to compute conventional results.

MIKE for an isolated system, Eq. (63), differs from Eq. (1) by the appearance of $i$, a result of taking the Laplace instead of Fourier transform. In this paper, we have used the former for at least three reasons. First, we wanted to avoid imaginary masses, such as arise, for example, in Eqs. (2.1) and (2.2) of [13]. (At the expense of adding an extra factor $\theta(m^2)$ in the Klein-Gordon equation, this issue is circumvented in Eq. (1) by taking the Fourier transform of $\theta(m^2)\phi$ instead of $\phi$.) Second, we have derived some interesting relations involving the convolution; but the inverse Fourier transform of the product of a function and its complex conjugate is a correlation not a convolution. Third, we have made use of the property that the Laplace transform of a positive kernel is another positive kernel, a result which is not generally true for the Fourier transform. These and other properties of MIKE make it worthy of study.

### VII. Appendix: The Manifold of Positive Cone Solutions

In the solutions $e^{-ikx}$ of the Klein-Gordon equation, $k$ is not any four-vector with $k^0 > 0$, but specific ones where $k^2 = m^2 \equiv \xi$ with fixed mass. In this appendix we would like to relax this last condition so that mass can vary in the solutions $e^{-i(\omega_{k,\xi} x^0 - \mathbf{k} \cdot \mathbf{x})}$, as in [4], [5] and [10], for example. The four-vector $k$ will be timelike (i.e., $k^2 > 0$) with positive energy, but $k^2$ will be able to assume any positive value. For such $k$, linear combinations of $e^{-ikx}$ will be called positive cone solutions, since a timelike four-vector $k$ with $k^0 > 0$ lies inside the three dimensional conical surface, $k^2 = 0$, in the upper half of the plane $k^0 = 0$.

For positive cone solutions, we may proceed analogously to the manner in which the manifold of positive energy solutions is obtained. Thus, we briefly recall some of the germane steps in the relativistic quantum mechanics of the Klein-Gordon particle as appears, for example, in [12].

An arbitrary eigenket $|k\rangle$ of the four-momentum operator $\hat{k}$ is not generally in the kernel of the Klein-Gordon operator because the dispersion relation may not hold. In other words, for arbitrary $k$ and fixed $m$, we generally have $k^2 - m^2 \neq 0$ and consequently $(\hat{k}^2 - m^2)|k\rangle \neq 0$. Thus, we consider kets $|\mathbf{k}\rangle$ that have positive energies and satisfy the dispersion relation, which can be achieved by choosing $k^0 = \omega_\mathbf{k} = \sqrt{\mathbf{k}^2 + m^2}$, and that are orthogonal according to

$$\langle \mathbf{k'}|\mathbf{k}\rangle = \alpha(\mathbf{k})\delta(\mathbf{k} - \mathbf{k'}) \tag{130}$$

for some function $\alpha(\mathbf{k})$. For the resolution of the identity for single particle states, it seems reasonable to choose

$$\int d^4 k |\mathbf{k}\rangle \frac{\delta(k^2 - m^2)\theta(k^0)}{(2\pi)^3} \langle \mathbf{k}| = 1_{KG} \tag{131}$$

or, in another form,

$$\int d^3 k |\mathbf{k}\rangle \frac{1}{(2\pi)^3 2\omega_\mathbf{k}} \langle \mathbf{k}| = 1_{KG}, \tag{132}$$

which in addition to having the dispersion relation and the positive energy restriction baked in, also has the virtue of being Lorentz invariant. (The $(2\pi)^3$ factor is chosen to conform to convention and $\delta(k^2 - m^2)\theta(k^0)d^4 k \to \frac{1}{2\omega_\mathbf{k}} d^3 k$ was used--see [11], Eq. 4.4) For Eqs. (130) and (132) to be true, we then require that $\alpha(\mathbf{k}) = (2\pi)^3 2\omega_\mathbf{k}$. We define a ket $|x\rangle$ such that $\langle x|\mathbf{k}\rangle = e^{-i(\omega_\mathbf{k} x^0 - \mathbf{k}\cdot\mathbf{x})}$ and then derive using Eq. (132)

$$|x\rangle = \int d^3 k \frac{1}{(2\pi)^3 2\omega_\mathbf{k}} |\mathbf{k}\rangle e^{i(\omega_\mathbf{k} x^0 - \mathbf{k}\cdot\mathbf{x})}. \tag{133}$$

The positive energy solutions are then

$$\left\langle x | \phi_{KG}^{(+)} \right\rangle \equiv \phi_{KG}^{(+)}(x)$$
$$= \int d^3 k \frac{1}{(2\pi)^3 2\omega_\mathbf{k}} e^{-i(\omega_\mathbf{k} x^0 - \mathbf{k}\cdot\mathbf{x})} \Phi_{KG}^{(+)}(\mathbf{k}) \tag{134}$$

where $\Phi_{KG}^{(+)}(\mathbf{k}) = \left\langle \mathbf{k} | \phi_{KG}^{(+)} \right\rangle$. The general inner product between two positive energy solutions is

$$\left\langle \phi_{KG}^{(+)} | \psi_{KG}^{(+)} \right\rangle = \int d^3 k \, \Phi_{KG}^{(+)*}(\mathbf{k}) \frac{1}{(2\pi)^3 2\omega_\mathbf{k}} \Psi_{KG}^{(+)}(\mathbf{k}). \tag{135}$$

We now turn to positive cone solutions and consider eigenkets $|k\rangle$ of the four-momentum operator $\hat{k}$ that have positive energies ($k^0 > 0$) and also belong to the kernel of $\hat{k}^2 - m^2$, but this time with variable $m > 0$. Thus, we deal with timelike $k$ with $k^0 > 0$. Proceeding as above, we attempt to find a resolution of the identity for single particle states given by

$$\int d^4k |k\rangle \theta(k^2)\theta(k^0)\langle k| = 1_{\text{PC}} \tag{136}$$

where $\langle k'|k\rangle = \delta^4(k-k')$. These two requirements are consistent because the restriction of $\int d^4k |k\rangle\langle k|$ to the positive cone solutions is equal to the left-hand side of Eq. (136). To see this, note that

$$\int d^4k |k\rangle \theta(k^2)\theta(k^0)\langle k| = \int d^4k |k\rangle\langle k| - \int d^3k \int_{-\infty}^{\|\mathbf{k}\|} dk^0 |k\rangle\langle k|. \tag{137}$$

The last term of Eq. (137) represents an operator obtained by integrating energy over the intervals $(-\infty, -\|\mathbf{k}\|)$, $(-\|\mathbf{k}\|, 0)$ and $(0, \|\mathbf{k}\|)$ associated respectively with timelike with negative energy, spacelike with negative energy and spacelike with positive energy four-vectors; when operating on a positive cone solution, a linear combination of timelike four-vectors with positive energy, this operator yields zero.

We now change variables from $(k^0, \mathbf{k})$ to $(\xi, \mathbf{k})$ according to $k^0 = \sqrt{\mathbf{k}^2 + \xi} \equiv \omega_{\mathbf{k},\xi}$ to rewrite Eq. (136) as

$$\int_0^\infty d\xi \int d^3k |\omega_{\mathbf{k},\xi}, \mathbf{k}\rangle \frac{1}{2\omega_{\mathbf{k},\xi}} \langle \omega_{\mathbf{k},\xi}, \mathbf{k}| = 1_{\text{PC}}, \tag{138}$$

which is Eq. (30). In view of Eq. (132), Eq. (138) appears to be $(2\pi)^3$ times the sum, over all masses, of the projections onto the positive energy solutions. (In geometric terms, this is related to the fact that the union over all $\xi > 0$ of the hyperboloids $k^2 = \xi$, with $k^0 > 0$, yields the interior of a three-dimensional conical surface.) However, note that in Eqs. (132) and (138), we have the following different inner products, the conventional one $\langle \mathbf{k}|\mathbf{k}'\rangle = (2\pi)^3 2\omega_{\mathbf{k}} \delta(\mathbf{k} -$

**k**′),

and

$$\langle\omega_{\mathbf{k},\xi},\mathbf{k}|\omega_{\mathbf{k}',\xi'},\mathbf{k}'\rangle = \delta(\omega_{\mathbf{k},\xi} - \omega_{\mathbf{k}',\xi'})\delta(\mathbf{k}-\mathbf{k}')$$
$$= 2\omega_{\mathbf{k},\xi}\delta(\xi-\xi')\delta(\mathbf{k}-\mathbf{k}') \quad (139)$$

the last expression being the same as the right-hand side of Eq. (5), apart from the inclusion of $r_0$.

We can confirm the validity of Eq. (138) on the manifold of states $\{|\omega_{\mathbf{k},\xi},\mathbf{k}\rangle\}$:

$$\int d^4k\,|k\rangle\theta(k^2)\theta(k^0)\langle k|\omega_{\mathbf{k}'},\mathbf{k}'\rangle$$
$$= \int d^4k\,|k\rangle\theta(k^2)\theta(k^0)\delta(k^0 - \omega_{\mathbf{k}',\xi'})\delta(\mathbf{k}-\mathbf{k}') \quad (140)$$
$$= |\omega_{\mathbf{k}',\xi'},\mathbf{k}'\rangle$$

We now define a ket $|x\rangle$ such that $\langle x|\omega_{\mathbf{k},\xi},\mathbf{k}\rangle = e^{-i(\omega_{\mathbf{k},\xi}x^0 - \mathbf{k}\cdot\mathbf{x})}$ and using Eq. (138) find

$$|x\rangle = \int_0^\infty d\xi \int d^3k\,|\omega_{\mathbf{k},\xi},\mathbf{k}\rangle \frac{1}{2\omega_{\mathbf{k},\xi}} e^{i(\omega_{\mathbf{k},\xi}x^0 - \mathbf{k}\cdot\mathbf{x})} \quad (141)$$

from which we obtain the positive cone solutions

$$\langle x|\phi_{PC}^{(+)}\rangle = \phi_{PC}^{(+)}(x)$$
$$= \int_0^\infty d\xi \int d^3k \frac{1}{2\omega_{\mathbf{k},\xi}} e^{-i(\omega_{\mathbf{k},\xi}x^0 - \mathbf{k}\cdot\mathbf{x})} \Phi_{PC}^{(+)}(\omega_{\mathbf{k},\xi},\mathbf{k}) \quad (142)$$

where

$\Phi_{PC}^{(+)}(\omega_{\mathbf{k},\xi},\mathbf{k}) = \langle\omega_{\mathbf{k},\xi},\mathbf{k}|\phi_{PC}^{(+)}\rangle$. These are linear combinations of $e^{-i(\omega_{\mathbf{k},\xi}x^0 - \mathbf{k}\cdot\mathbf{x})}$, or in ket form of $|\omega_{\mathbf{k},\xi},\mathbf{k}\rangle$, in which the sums are over both $\mathbf{k}$ and $\xi$.

In the conventional theory, it is well known [12] that the operator $\hat{\mathbf{X}}_{KG}$, defined in momentum space by $i\boldsymbol{\nabla}_{\mathbf{k}}$, is not self-adjoint, so that the following expression is not equal to zero:

$$\langle \mathbf{k}'|\hat{\mathbf{X}}_{KG}|\mathbf{k}\rangle - \langle \mathbf{k}|\hat{\mathbf{X}}_{KG}|\mathbf{k}'\rangle^* \\ = i(2\pi)^3 2[\omega_{\mathbf{k}}\boldsymbol{\nabla}_{\mathbf{k}'}\delta(\mathbf{k}'-\mathbf{k}) + \omega_{\mathbf{k}'}\boldsymbol{\nabla}_{\mathbf{k}}\delta(\mathbf{k}-\mathbf{k}')]. \tag{143}$$

In contrast, the position operator $\hat{X}^\mu_{PC}$ for positive cone solutions, defined in momentum space by $-i\frac{\partial}{\partial k_\mu}$, is self-adjoint. To wit,

$$\langle k'|\hat{X}^\mu_{PC}|k\rangle - \langle k|\hat{X}^\mu_{PC}|k'\rangle^* = -i\frac{\partial \delta^4(k'-k)}{\partial k'_\mu} - i\frac{\partial \delta^4(k-k')}{\partial k_\mu} \tag{144}$$
$$= 0.$$

Another resolution of the identity is

$$\frac{1}{(2\pi)^4}\int d^4x |x\rangle\langle x| = \mathbf{1}_{PC}, \tag{145}$$

which may be verified as follows:

$$\langle \omega_{\mathbf{k}',\xi'}, \mathbf{k}'| \frac{1}{(2\pi)^4}\int d^4x |x\rangle\langle x|\omega_{\mathbf{k},\xi}, \mathbf{k}\rangle \\ = \frac{1}{(2\pi)^4}\int d^4x\, e^{i(\omega_{\mathbf{k}',\xi'}x^0 - \mathbf{k}'\cdot\mathbf{x})} e^{-i(\omega_{\mathbf{k},\xi}x^0 - \mathbf{k}\cdot\mathbf{x})} \tag{146}$$
$$= \delta(\omega_{\mathbf{k}',\xi'} - \omega_{\mathbf{k},\xi})\delta(\mathbf{k}-\mathbf{k}').$$

Using Eq. (138), we may compute the inner product of two positive cone solutions thusly:

$$\langle \phi^{(+)}_{PC}|\psi^{(+)}_{PC}\rangle = \int_0^\infty d\xi \int d^3k\, \Phi^{(+)*}_{PC}(\omega_{\mathbf{k},\xi}, \mathbf{k}) \frac{1}{2\omega_{\mathbf{k},\xi}} \Psi^{(+)}_{PC}(\omega_{\mathbf{k},\xi}, \mathbf{k}) \tag{147}$$

The positive cone solutions may also be written as

$$\phi^{(+)}_{PC}(x) = \int d^4k\, e^{-ikx}\theta(k^2)\theta(k^0)\Phi^{(+)}_{PC}(k) \tag{148}$$

and in anticipation of introducing charged particles in quantum field theory, we also define

$$\phi_{PC}^{(-)}(x) = \int d^4k \, e^{ikx} \theta(k^2)\theta(k^0)\Phi_{PC}^{(-)}(k) \tag{149}$$

and

$$\phi_{PC}(x) = \phi_{PC}^{(+)}(x) + \phi_{PC}^{(-)}(x). \tag{150}$$

We may transition to quantum field theory in the usual way by replacing $\Phi_{PC}^{(+)}(\omega_{\mathbf{k},\xi}, \mathbf{k})$ with $a(\omega_{\mathbf{k},\xi}, \mathbf{k})$ and $\Phi_{PC}^{(-)}(\omega_{\mathbf{k},\xi}, \mathbf{k})$ with $b^\dagger(\omega_{\mathbf{k},\xi}, \mathbf{k})$, where $a^\dagger(k)|0;0\rangle = |k;0\rangle$, $b^\dagger(k)|0;0\rangle = |0;k\rangle$, and $[a(k), a^\dagger(k')] = [b(k), b^\dagger(k')] = \delta^4(k - k')$; these commutators appear in the literature [3]. We then obtain the field operators

$$\phi_{PC}^{(+)}(x) = \int d^4k \, e^{-ikx} \theta(k^2)\theta(k^0) a(k), \tag{151}$$

$$\phi_{PC}^{(-)}(x) = \int d^4k \, e^{ikx} \theta(k^2)\theta(k^0) b^\dagger(k) \tag{152}$$

and

$$\phi_{PC}(x) = \phi_{PC}^{(+)}(x) + \phi_{PC}^{(-)}(x). \tag{153}$$

For particles, as opposed to antiparticles, we obtain

$$|x\rangle = \phi_{PC}^\dagger(x)|0\rangle. \tag{154}$$

The field operators as just defined are independent of $\xi$, since we integrate out to infinite mass; when we then compute the following fundamental commutator, we obtain

$$[\phi_{PC}(t,\mathbf{x}), \dot{\phi}^\dagger_{PC}(t,\mathbf{y})] = i(2\pi)^3 \delta(\mathbf{x}-\mathbf{y}) \int_0^\infty d\xi, \tag{155}$$

which diverges linearly in $\xi$. To remedy this, we may introduce a high energy cutoff $k^0_{max} = \sqrt{\mathbf{k}^2 + \xi_{max}}$; the mass cutoff $\xi_{max}$ can coincide with $\xi_p$ for example, the particular mass of interest. Let us define

$$\phi^{(+)}_{PC}(x;\xi_{max}) = \int_0^{k^0_{max}} dk^0 \int d^3k\, e^{-ikx} \theta(k^2)\theta(k^0) a(k), \tag{156}$$

$$\phi^{(-)}_{PC}(x;\xi_{max}) = \int_0^{k^0_{max}} dk^0 \int d^3k\, e^{ikx} \theta(k^2)\theta(k^0) b^\dagger(k) \tag{157}$$

and

$$\phi_{PC}(x;\xi_{max}) = \phi^{(+)}_{PC}(x;\xi_{max}) + \phi^{(-)}_{PC}(x;\xi_{max}) \tag{158}$$

We then have

$$[\phi_{PC}(t,\mathbf{x};\xi_{max}), \xi^{-1}_{max}\dot{\phi}^\dagger_{PC}(t,\mathbf{y};\xi_{max})] = i(2\pi)^3 \delta(\mathbf{x}-\mathbf{y}). \tag{159}$$

Many of the relationships provided above without a mass cutoff continue to hold with a cutoff. For example, we have the following single particle resolution of the identity,

$$\int_0^{\xi_{max}} d\xi \int d^3k |\omega_{\mathbf{k},\xi}, \mathbf{k}\rangle \frac{1}{2\omega_{\mathbf{k},\xi}} \langle \omega_{\mathbf{k},\xi}, \mathbf{k}| = 1_{PC}, \tag{160}$$

on the set $\{|\omega_{\mathbf{k},\xi}, \mathbf{k}\rangle | 0 < \xi < \xi_{max}\}$.